\documentclass[12pt]{article}
\usepackage{amsmath}
\usepackage{amssymb}
\usepackage{multirow}
\usepackage{graphicx}
\usepackage[dvips]{color}
\usepackage{ulem}
\makeatletter\@addtoreset{equation}{section}
\makeatother

\begin{document}

\begin{titlepage}
\begin{flushright}
TIT/HEP-654 \\
May,  2016
\end{flushright}
\vspace{0.5cm}
\begin{center}
{\Large \bf
 ODE/IM correspondence for modified $B_2^{(1)}$ 
affine Toda field equation 
}
\lineskip .75em
\vskip 2.5cm
{\large  Katsushi Ito and Hongfei Shu }
\vskip 2.5em
 {\normalsize\it Department of Physics,\\
Tokyo Institute of Technology\\
Tokyo, 152-8551, Japan} 
\vskip 3.0em
\end{center}
\begin{abstract}        
We study the massive ODE/IM correspondence for modified $B_{2}^{(1)}$
affine Toda field equation. Based on the $\psi $-system for the
solutions of the associated linear problem, we obtain the Bethe ansatz
equations. We also discuss the T--Q relations, the T-system and the
Y-system, which are shown to be related to those of the $A_{3}/{\mathbf{Z}}
_{2}$ integrable system. We consider the case that the solution of the
linear problem has a monodromy around the origin, which imposes
nontrivial boundary conditions for the T-/Y-system. The high-temperature
limit of the T- and Y-system and their monodromy dependence are studied
numerically.
\end{abstract}
\end{titlepage}
%
%


\section{Introduction}

It has been recognized that the relation between classical and quantum
integrable systems is useful for studying non-perturbative properties
of supersymmetric gauge theories and the AdS/CFT correspondence
\cite{Gaiotto:2009hg,Nekrasov:2009rc,Alday:2009dv}. The ODE/IM
correspondence \cite{Dorey:1998pt,Bazhanov:1998wj,Dorey:2007zx}
provides an interesting example of this classical/quantum
correspondence, which relates the spectral determinants of certain
ordinary differential equations (ODE) to the Bethe ansatz equations in the
massless limit of certain integrable models (IM). It is an interesting
problem to make a complete list of this ODE/IM correspondence. The
ordinary differential equations for the integrable models related to
classical Lie algebras have been proposed in \cite{Dorey:2006an}. The
Wronskian of the solutions obeys the functional relations called the
$\psi $-system, which leads to the Bethe ansatz equations of the related
quantum integrable system. The $\psi $-system for classical Lie algebra
has been reformulated in the form of the matrix valued linear
differential equations \cite{juanjuan2012polynomial}, where the Bethe
ansatz equations of the integrable models associated with the untwisted
affine Lie algebra $X^{(1)}$ of a classical Lie algebra $X$ are related
to the linear differential equations associated with the Langlands dual
$(X^{(1)})^{\vee }$.

The ODE/IM correspondence has been generalized to massive integrable
models. It was found that for the classical sinh-Gordon equation
modified by a conformal transformation, the spectral problem for the
associated linear problem leads to the functional relations of the
quantum sine-Gordon model \cite{Lukyanov:2010rn}. By taking the
conformal limit, it reduces to the ODE/IM correspondence for the
Schr\"{o}dinger type differential equation
\cite{Dorey:1998pt,Bazhanov:1998wj}.

Recently the massive ODE/IM correspondence has been generalized to a
class of modified affine Toda field equations
\cite{Dorey:2012bx,Ito:2013aea,Adamopoulou:2014fca,Ito:2015nla,Masoero:2015lga,Masoero:2015rcz}.
In particular, Locke and one of the present authors studied the modified
affine Toda equations for affine Lie algebra $\hat{\mathfrak{g}}^{
\vee }$, where $\hat{\mathfrak{g}}$ is an untwisted affine Lie algebra
including exceptional type \cite{Ito:2015nla}. It has been shown that
from their associated linear problems one obtains the $\psi $-system
which leads to the Bethe ansatz equations for the affine Lie algebra
$\hat{\mathfrak{g}}$.

It would be an interesting problem to explore the modified affine Toda
field equation with an affine Lie algebra $\hat{\mathfrak{g}}$ which is
not of the form of the Langlands dual of an untwisted one, where the
corresponding integrable models are not identified yet. In this paper
we will work with the modified affine Toda field equation associated
with the affine Lie algebra $B_{2}^{(1)}$ (or $C_{2}^{(1)}$), which
provides the simplest and nontrivial example. This equation also appears
in the study of the area of minimal surface with a null-polygonal
boundary in $AdS_{4}$ spacetime
\cite{DeVega:1992xc,Burrington:2009bh,Burrington:2011eh,Alday:2010vh},
which is dual to the gluon scattering amplitudes with specific momentum
configurations. The equation of motion of strings is described by the
$B_{2}^{(1)}$ affine Toda field equation modified by the conformal
transformation. The Stokes problem of the associated linear system
determines the functional equations for the cross-rations of external
momenta. These functional relations are known to be the same as the
Y-system of the homogeneous sine-Gordon model
\cite{Hatsuda:2010cc,Hatsuda:2011ke} and the free energy of the Y-system
determines the area of the minimal surface.

The purpose of this paper is to apply the massive ODE/IM correspondence
to the modified $B_{2}^{(1)}$ affine Toda field equation and to
investigate the functional relations for the Stokes coefficients of the
linear problem, which include the Bethe ansatz equations, the T--Q
relations, the T-system and the Y-system. We study the boundary
condition of the T-system arising from the nontrivial monodromy of the
linear problem solution around the origin. This monodromy condition also
appears in the study of the form factors via the AdS/CFT correspondence
\cite{Maldacena:2010kp,Gao:2013dza}.

This paper is organized as follows: In sect.~\ref{sec2}, we introduce modified
$B_{2}^{(1)}$ affine Toda equation and the associated linear problem.
In sect.~\ref{sec3}, we discuss the $\psi $-system and derive the Bethe ansatz
equations. In sect.~\ref{sec4}, we discuss the spinor representation of
$B_{2}$ in detail and study the quantum Wronskian and the T--Q relations.
In sect.~\ref{sec5}, we argue the T-system and Y-system and their boundary
conditions which come from the monodromy of the solution of the linear
system around the origin. In sect.~\ref{sec6}, we investigate the
high-temperature limit of the Y-system in the presence of monodromy.
Sect.~\ref{sec7} is devoted for conclusions and discussion. In the appendix, we
summarize the auxiliary T-functions and their functional relations used
in this paper.\looseness=1

\section{Modified $B_{2}^{(1)}$ affine Toda field equation}
\label{sec2}
The Lie algebra $B_{2}=so(5)$ has simple roots $\alpha_{1}=e_{1}-e
_{2}$ and $\alpha_{2}=e_{2}$, where $e_{i}$ ($i=1,2$) is an orthonormal
basis of $\mathbf{R}^{2}$. We denote the highest root by $\theta $, which
is given by $\theta =\alpha_{1}+2\alpha_{2}$, and define the extended
root $\alpha_{0}=-\theta $. $\omega_{1}=e_{1}$ and $\omega_{2}=
\frac{1}{2}(e_{1}+e_{2})$ are the fundamental weights satisfying
$2\omega_{i}\cdot \alpha_{j}/\alpha_{j}^{2}=\delta_{ij}$. Let
$\{H^{i},E_{\alpha }\}$ ($i=1,2$, $\alpha \in \Delta $ ) be the
Chevalley basis of $B_{2}$, where $\Delta $ is the set of roots.

Let $\phi =(\phi^{1},\phi^{2})$ be the two-component scalar field on the
complex plane with coordinates $(z,\bar{z})$. We define the modified
affine Toda field equation for $B_{2}^{(1)}$ by
\begin{equation}
\partial \bar{\partial } \phi -\frac{m^{2}}{\beta }
\left(
\alpha_{1} e
^{\beta \alpha_{1}\cdot \phi }+2\alpha_{2}e^{\beta \alpha_{2}\cdot
\phi }
+p(z)\bar{p}(\bar{z}) \alpha_{0} e^{\beta \alpha_{0}\cdot
\phi }
\right) =0.
\label{eq:mafe1}
\end{equation}
where $\partial =\frac{\partial }{\partial z}$, $\bar{\partial }=
\frac{\partial }{\partial \bar{z}}$, $m$ is a mass parameter and
$\beta $ is a coupling parameter. $p(z)$ is a holomorphic function of $z$ and is
chosen as
\begin{equation}
p(z)=z^{4M}-s^{4M},
\label{eq:pz1}
\end{equation}
with $M>\frac{1}{3}$ and $s$ is a complex parameter. This equation is
obtained by the conformal transformation $z\rightarrow w$ with
$\frac{\partial w}{\partial z}=p^{\frac{1}{4}}$ and the field redefinition
$ \phi \rightarrow \phi -\frac{1}{4\beta }\rho^{\vee }\log (p\bar{p})$,
where $\rho^{\vee }=\omega_{1}+2\omega_{2}$ is the co-Weyl
vector.\footnote{Note that the modified equation in
\cite{Burrington:2009bh} is $\partial \bar{\partial }\phi -
\frac{m^{2}}{\beta }(\alpha_{1}\sqrt{p\bar{p}}e^{\beta \alpha_{1}
\cdot \phi }+2\alpha_{2}e^{\beta \alpha_{2}\cdot \phi }+\sqrt{p
\bar{p}}\alpha_{0}e^{\beta \alpha_{0}\cdot \phi })=0$, which is obtained
by the same conformal transformation but a different field redefinition
$ \phi \rightarrow \phi -\frac{1}{4\beta }\alpha_{2}\log (p\bar{p})$. This
modified equation is related to (\ref{eq:mafe1}) by a field
redefinition. } Note that the Coxeter number of $B_{2}$ is 4. Eq.~(\ref{eq:mafe1})
can be written in the form of the compatibility
condition $[\partial +A_{z}, \bar{\partial }+A_{\bar{z}}]=0$ of the
linear differential equations defined in a $B_{2}$-module:
\begin{equation}
(\partial +A_{z})\Psi =0,\quad (\bar{\partial }+\bar{A}_{\bar{z}})
\Psi =0,
\label{eq:linear1}
\end{equation}
where the connections are defined by
\begin{align}
A_{z}&=\frac{\beta }{2}\partial \phi \cdot H+m e^{\lambda }
\left( e^{
\beta \alpha_{1}\cdot \phi /2}E_{\alpha_{1}}
+e^{\beta \alpha_{2}
\cdot \phi /2}E_{\alpha_{2}}
+p(z)e^{\beta \alpha_{0}\cdot \phi /2}E
_{\alpha_{0}}
\right) ,
\nonumber
\\
\bar{A}_{\bar{z}}&=-\frac{\beta }{2}\bar{\partial }\phi \cdot H+m e^{-
\lambda }
\left( e^{\beta \alpha_{1}\cdot \phi /2}E_{-\alpha_{1}}
+e
^{\beta \alpha_{2}\cdot \phi /2}E_{-\alpha_{2}}
+\bar{p}(\bar{z})e
^{\beta \alpha_{0}\cdot \phi /2}E_{-\alpha_{0}}
\right) .
\label{eq:conn1}
\end{align}
Here $\lambda $ is the spectral parameter. We are interested in the
special class of solutions of (\ref{eq:mafe1}), which satisfy the
periodicity condition $\phi (\rho ,\theta +\frac{\pi }{4M})=\phi (
\rho ,\theta )$ and the boundary conditions at infinity and the origin
of the complex plane:
\begin{align}
\phi (\rho ,\theta )
&= \frac{2M\rho ^{\vee } }{\beta }\log \rho +
\cdots , \quad (\rho \rightarrow \infty )\\
\phi (\rho ,\theta )
&= 2g \log \rho +\cdots , \quad (\rho \rightarrow
0)
\end{align}
where we have introduced the polar coordinate $(\rho ,\theta )$ by
$z=\rho e^{i\theta }$ and $g$ is a 2-vector satisfying $\beta
\alpha_{a}\cdot g+1>0$ ($a=0,1,2$). Due to the special form
(\ref{eq:pz1}) of $p(z)$, (\ref{eq:mafe1}) and the linear problem are
invariant under the Symanzik rotation
\begin{align}
\hat{\Omega }_{k} : (z,s,\lambda )\rightarrow (z e^{
\frac{2\pi i k}{4M}}, s e^{\frac{2\pi i k}{4M}}, \lambda -
\frac{2\pi i k}{4M}),
\end{align}
for an integer $k$. This also acts on the solution $\Psi (z,\bar{z})$,
which is denoted as $\Psi_{k}(z,\bar{z}):=\hat{\Omega }_{k}\Psi (z,
\bar{z})$. The linear problem is also invariant under the
transformation:
\begin{align}
\hat{\Pi } : (\lambda , A_{z}, A_{\bar{z}},\Psi )
\rightarrow
(\lambda
-\frac{2\pi i}{4}, SA_{z} S^{-1}, SA_{\bar{z}}S^{-1}, S\Psi )
\end{align}
where $S=\exp (\frac{2\pi i }{4}\rho^{\vee }\cdot H)$.

We now consider the solutions of the linear differential equations
(\ref{eq:linear1}) in the basic $B_{2}^{(1)}$-module $V^{(a)}$ ($a=1,2$)
associated with the highest weight $\omega_{a}$. Let $\mathbf{e}_{j}^{(a)}$
be the orthonormal basis of $V^{(a)}$ with $H^{i}$ eigenvalue
$(h_{j}^{(a)})^{i}$, where $i,j=1,\cdots , \text{dim}V^{(a)}$. For the
Lie algebra $B_{2}$, $V^{(1)}$ is 5-dimensional vector representation,
whose matrix representation is given by
\begin{align}
E_{\alpha_{1}}
&=e_{1,2}+e_{4,5},\quad
E_{\alpha_{2}}=\sqrt{2}(e_{2,3}+e
_{3,4}),\quad
E_{\alpha_{0}}=-(e_{4,1}+e_{5,2})
\label{eq:vectrep1}
\end{align}
and $E_{-\alpha_{i}}=E^{T}_{\alpha_{i}}$. Here $e_{ab}$ denotes the
matrix whose $(i,j)$-element is $\delta_{i a}\delta_{j b}$. Similarly,
$V^{(2)}$ is a 4-dimensional spinor representation. Its matrix
representation is given by
\begin{align}
E_{\alpha_{1}}
&=e_{23}, \quad E_{\alpha_{2}}=e_{12}+e_{34}, \quad
E
_{\alpha_{0}}=e_{41}
\label{eq:spinrep1}
\end{align}
and $E_{-\alpha_{i}}=E_{\alpha_{i}}^{T}$.

We are interested in the small (or subdominant) solution $\Psi^{(a)}$,
which decays fastest along the positive real axis. This was studied in
\cite{Ito:2015nla} for $\hat{\mathfrak{g}}^{\vee }$ for an untwisted
affine Lie algebra ${\mathfrak{g}}$. In general, the small solution
$\Psi^{(a)}$ at large $\rho $ is given by
\begin{align}
\Psi^{(a)}(z,\bar{z}|\lambda ,g)
&=
C^{(a)} \exp \left(
-2\mu^{(a)}
\frac{\rho ^{M+1}}{M+1} m \cosh (\lambda +i\theta (M+1))
\right)
e^{-i
\theta M\rho^{\vee }\cdot H}{\boldsymbol{\mu }}^{(a)},
\label{eq:subdom1}
\end{align}
with $C^{(a)}$ being a normalization constant. Here ${\boldsymbol{\mu }}
^{(a)}$ and $\mu^{(a)}$ denote the eigenvector and its eigenvalue of the
matrix $\Lambda_{+}=E_{\alpha_{0}}+E_{\alpha_{1}}+E_{\alpha_{2}}$ with
the eigenvalue of the largest real part. Applying the Symanzik rotation
$\hat{\Omega }_{k}$ ($k\in {\mathbf{Z}}$), one obtains the small solution
$\Psi_{k}^{(a)}$ in the Stokes sector
\begin{align}
{\mathcal{S}}_{-k}:
\left|  \theta +\frac{2\pi k}{4(M+1)}\right| <
\frac{\pi }{4(M+1)}
\end{align}

For the vector representation (\ref{eq:vectrep1}), the eigenvalues of
$\Lambda_{+}$ are $\sqrt{2}e^{\frac{i\pi }{4}(2k+1)}$ ($k=0,1,2,3$) and~$0$.
For the spinor representation (\ref{eq:spinrep1}), they are
$\pm 1$ and $\pm i$. For $V^{(1)}$, one has two eigenvalues with the
largest real part and the corresponding solutions in $V^{(1)}$ are not
subdominant along the real axis. So we introduce the $\frac{1}{2}$-rotated
Symanzik solution $\Psi_{\frac{1}{2}}^{(1)}$. This is a solution of the
linear problem with the $\frac{1}{2}$-rotated connection
$(A_{\frac{1}{2}})_{z}$ and $(\bar{A}_{\frac{1}{2}})_{\bar{z}}$ which is
obtained by replacing $E_{\pm \alpha_{0}}\rightarrow -E_{\pm \alpha
_{0}}$ in (\ref{eq:conn1}). Then the $\Psi_{\frac{1}{2}}^{(1)}$ behaves
along the real positive axis as (\ref{eq:subdom1}) with $\mu^{(1)}=
\sqrt{2}$ and ${\boldsymbol{\mu }}^{(1)}=(1,\sqrt{2},\sqrt{2},
\sqrt{2},1)^{T}$.

We define the basis of the solutions around $\rho =0$ behaves as
$\rho \rightarrow 0$:
\begin{align}
{\mathcal{X}}_{i}^{(a)}(z,\bar{z}|\lambda ,g)
&=e^{-(\lambda +i\theta )
\beta g\cdot h_{i}^{(a)}}
{\mathbf{e}}_{i}^{(a)}+O(\rho ),
\quad
i=1,\cdots , \text{dim}V^{(a)}
\end{align}
which are invariant under $\hat{\Omega }_{k}$ \cite{Ito:2015nla}. The
small solution $\Psi^{(1)}_{\frac{1}{2}}$ and $\Psi^{(2)}$ can be expanded
in this basis as
\begin{align}
\Psi^{(1)}_{\frac{1}{2}}(z,\bar{z}|\lambda ,g)
&=\sum_{i=1}^{5}
Q^{(1)}
_{i}(\lambda ,g)\mathcal{X}_{i}^{(1)}(z,\bar{z}|\lambda ,g),
\nonumber
\\
\Psi^{(2)}(z,\bar{z}|\lambda ,g)
&=\sum_{i=1}^{4}
Q^{(2)}_{i}(\lambda
,g)\mathcal{X}_{i}^{(2)}(z,\bar{z}|\lambda ,g).
\label{eq:qfunct1}
\end{align}
We call $Q_{i}^{(a)}(\lambda ,g)$ the Q-functions. From the relation
$\hat{\Omega }_{1}\hat{\Pi }\Psi^{(a)}=\Psi^{(a)}$, the coefficients
$Q^{(a)}_{i}(\lambda ,g)$ satisfy the quasi-periodicity condition:
\begin{align}
Q_{i}^{(a)}(\lambda -\frac{2\pi i}{4M}(M+1),g)
=\exp (-\frac{2\pi i}{4}(
\rho^{\vee }+\beta g)\cdot h_{i}^{(a)})
Q_{i}^{(a)}(\lambda ,g).
\label{eq:period1}
\end{align}
Note that we can rescale $z$ and $\bar{z}$ such that the mass parameter
$m$ is fixed to be an arbitrary non-zero constant. Then the Q-functions
depend on the mass parameter through $s/m$.

\section{$\psi $-System and the Bethe ansatz equations}
\label{sec3}
The linear problem in the basic $B_{2}^{(1)}$-modules $V^{(a)}$ can be
also defined in other $B_{2}$-modules corresponding to the
(anti-)symmetrized tensor product of $V^{(a)}$'s. The inclusion maps
between the modules induce the relation between the small solutions,
which is called the $\psi $-system \cite{Dorey:2006an}. For example,
we consider the inclusion map
\begin{align}
&\iota_{1}
 : V^{(1)}\wedge V^{(1)} \hookrightarrow V^{(2)}\otimes V
^{(2)},
\\
&\iota_{2}
 : V^{(2)}\wedge V^{(2)} \hookrightarrow V^{(1)}.
\end{align}
By these maps the highest weight state $\mathbf{e}^{(1)}_{1}\wedge {\mathbf{e}}
^{(1)}_{2}$ is mapped to $\sqrt{2}{\mathbf{e}}^{(2)}_{1}\otimes {\mathbf{e}}
^{(2)}_{1}$ and $\mathbf{e}^{(2)}_{1}\wedge {\mathbf{e}}^{(2)}_{2}$ to
$\mathbf{e}_{1}^{(1)}$. We use this map to relate the solutions of the
linear problem defined on the different modules. $\Psi^{(1)}_{1}
\wedge \Psi^{(1)}_{0}$ is a solution of the linear problem
(\ref{eq:linear1}) on $V^{(1)}\wedge V^{(1)}$ due to invariance of
(\ref{eq:linear1}) under the Symanzik rotation $\hat{\Omega }_{1}$. This
solution is mapped into the module $V^{(2)}\otimes V^{(2)}$
by~$\iota_{1}$. Now $\Psi^{(2)}\otimes \Psi^{(2)}$ is the unique solution
in $V^{(2)}\otimes V^{(2)}$ with the same asymptotic behavior at large
$\rho $. In a similar way we can identify $\Psi^{(2)}_{\frac{1}{2}}
\wedge \Psi^{(2)}_{-\frac{1}{2}}$ with $\Psi^{(1)}_{\frac{1}{2}}$. Thus we
obtain the $\psi $-system:
\begin{align}
\iota_{1}(\Psi^{(1)}_{1}\wedge \Psi^{(1)}_{0})
&=\Psi^{(2)}\otimes
\Psi^{(2)},
\label{eq:psi-system1}\\
\iota_{2}(\Psi^{(2)}_{\frac{1}{2}}\wedge \Psi^{(2)}_{-\frac{1}{2}})
&=
\Psi^{(1)}_{\frac{1}{2}}
\label{eq:psi-system2}
\end{align}

Expanding the small solutions in the basis $\{ \mathcal{X}^{(a)}_{i} \}$
and substituting them into the $\psi $-system, one obtains the
functional relation for the Q-functions $Q^{(a)}_{1}$ and $Q^{(a)}
_{2}$:
\begin{align}
Q^{(1)}_{1}(\lambda -\frac{2\pi i}{8M})Q^{(1)}_{2}(\lambda +
\frac{2\pi i}{8M})
-Q^{(1)}_{2}(\lambda -\frac{2\pi i}{8M})Q^{(1)}_{1}(
\lambda +\frac{2\pi i}{8M})
&=2Q^{(2)}_{1}(\lambda ) Q^{(2)}_{1}(
\lambda ),
\\
Q^{(2)}_{1}(\lambda -\frac{2\pi i}{8M})
Q^{(2)}_{2}(\lambda +
\frac{2\pi i}{8M})
-Q^{(2)}_{2}(\lambda -\frac{2\pi i}{8M})
Q^{(2)}_{1}(
\lambda +\frac{2\pi i}{8M})
&=Q^{(1)}_{1}(\lambda ).
\end{align}
Denoting the zeros of the Q-functions $Q^{(a)}_{1}(\lambda )$ by
$\lambda^{(a)}_{1 n}$ ($n=1,2,\ldots $), one obtains the Bethe ansatz
equations
\begin{align}
\frac{Q_{1}^{(2)}(\lambda ^{(1)}_{1n}-\frac{\pi i}{4M})^{2}
}{Q_{1}^{(2)}(\lambda ^{(1)}_{1n}+\frac{\pi i}{4M})^{2}}
\frac{Q^{(1)}_{1}(\lambda ^{(1)}_{1n}+\frac{\pi i}{2M})
}{Q^{(1)}_{1}(\lambda ^{(1)}_{1n}-\frac{\pi i}{2M})}
&=-1,
\label{eq:bae1a}\\
\frac{Q^{(2)}_{1}(\lambda ^{(2)}_{1n}-\frac{\pi i}{2M})
}{Q^{(2)}_{1}(\lambda ^{(2)}_{1n}+\frac{\pi i}{2M})}
\frac{Q^{(1)}_{1}(\lambda ^{(2)}_{1n}+\frac{\pi i}{4M})
}{Q^{(1)}_{1}(\lambda ^{(2)}_{1n}-\frac{\pi i}{4M})}
&=-1.
\label{eq:bae1b}
\end{align}
Note that these differ from those of the integrable model based on the
$U_{q}(A_{3}^{(2)})$
\cite{reshetikhin1987towards,kuniba_suzuki_1995}, which is expected
from the Langlands duality between $A_{3}^{(2)}$ and $B_{2}^{(1)}$. The
Bethe ansatz equations for $U_{q}(A_{3}^{(2)})$ do not include the
squared Q-functions. It would be interesting to study the solutions of
the Bethe ansatz equations (\ref{eq:bae1a}) and (\ref{eq:bae1b}) in the
conformal limit and explore the corresponding integrable model.

\section{Quantum Wronskian and T--Q relations}
\label{sec4}
\subsection{Spinor representation and discrete symmetries}
\label{sec4.1}
The $\psi $-system in the previous section has been obtained by
investigating the asymptotic solution in a single Stokes sector,
$\mathcal{S}_{0}$ for example. Now we consider the solutions of the linear
problem in the whole complex plane. We focus on $V^{(2)}$ because this
is the minimal dimensional representation and the solution in the vector
representation can be constructed via the inclusion map $\iota_{2}$.

Since we are considering a $SO(5)$ spinor, it is natural to introduce
the charge conjugation. Associated with the linear problem
{(\ref{eq:linear1})} in the spinor representation, we define the
transposed linear problem:
\begin{align}
(\partial -A^{T}_{z})\bar{\Psi }=0,\quad
(\bar{\partial } -A^{T}_{
\bar{z}})\bar{\Psi }=0.
\label{eq:lineart1}
\end{align}
The solution $\bar{\Psi }(z,\bar{z}|\lambda ,g)$ of these equations are
related to $\Psi (z,\bar{z}|\lambda ,g)$ by the charge conjugation:
\begin{align}
\bar{\Psi }(z,\bar{z}|\lambda ,g)=F\Psi (z,\bar{z}|\lambda ,g),
\label{eq:charge1}
\end{align}
where
\begin{align}
F=
\begin{pmatrix}
0 & 0 & 0 & 1
\\
0 & 0 & -1 & 0
\\
0 & 1 & 0 & 0
\\
-1 & 0 & 0 & 0
\end{pmatrix}
 .
\end{align}
This is a $\mathbf{Z}_{2}$ symmetry of the linear problem. Note that
$\bar{\bar{\Psi }}=-\Psi $.

One can define the inner product $\langle \bar{\Psi }, \Psi \rangle :=
\sum_{\alpha =1}^{4} \bar{\Psi }^{\alpha }\Psi^{\alpha }$ between
$\Psi =(\Psi^{\alpha })$ and $\bar{\Psi }=(\bar{\Psi }^{\alpha })$. The
inner product is independent of $z$ and $\bar{z}$ when $\Psi $ ($\bar{
\Psi }$) is a solution of the (transposed) linear problem. The Wronskian
of any four linearly independent solutions $\Psi_{i}$ ($i=1,2,3,4$)
\begin{align}
\langle \Psi_{1},\Psi_{2},\Psi_{3},\Psi_{4}\rangle
:=\mathrm{det}(\Psi
_{1},\Psi_{2},\Psi_{3},\Psi_{4}),
\end{align}
is also independent of $z$ and $\bar{z}$.

We define the $(-k)$-rotated solution $s_{k}:=\Psi^{(2)}_{-k}$ in the
module $V^{(2)}$. This is the subdominant solution in the Stokes sector
$\mathcal{S}_{k}$ but it gives a divergent solution in the sectors
$\mathcal{S}_{k-2}$ and $\mathcal{S}_{k+2}$. One can choose $\{s_{k-1},s
_{k},s_{k+1}, s_{k+2}\}$ as a basis of the solutions. We normalize the
solution $s_{k}$ such that
\begin{align}
\langle s_{k-1}, s_{k},s_{k+1}, s_{k+2}\rangle =1,
\label{eq:nor1}
\end{align}
by choosing the normalization constant $C^{(2)}$ in {(\ref{eq:subdom1})}
as $(-16)^{-\frac{1}{4}}$. From the asymptotic behavior of $s_{k}$ and
$\bar{s}_{k}$ at large $\rho $, we find $\langle \bar{s}_{k},s_{k}
\rangle =\langle \bar{s}_{k},s_{k\pm 1}\rangle =0$ and $\langle
\bar{s}_{k}, s_{k+2}\rangle =\frac{1}{16}$. Then from the condition
{(\ref{eq:nor1})} we find
\begin{align}
\bar{s}_{k}^{\alpha }=-\frac{1}{16}
\epsilon^{\alpha \beta_{1}\beta_{2}\beta_{3}}s_{k-1}^{\beta_{1}}s_{k}
^{\beta_{2}}s_{k+1}^{\beta_{3}}.
\label{eq:barsk2}
\end{align}
We write it in the form $\bar{s}_{k}=-\frac{1}{16}s_{k-1}\wedge s_{k}
\wedge s_{k+1}$. Since the basis $\mathbf{e}_{i}^{(2)}$ is orthonormal, we
can fix the normalization of $\mathcal{X}^{(2)}_{i}$ as
\begin{align}
{\mathrm{det}}(\mathcal{X}_{1}^{(2)}, \mathcal{X}_{2}^{(2)}, \mathcal{X}_{3}^{(2)},
\mathcal{X}_{4}^{(2)})=1,
\label{eq:normalization1}
\end{align}
which simplifies the functional relations described below.

\subsection{T--Q relation}

Now we take $\{s_{-2},s_{-1}, s_{0}, s_{1}\}$ as the basis of the
solutions of the linear system. We introduce a set of functions
$\mathcal{T}_{a,m}(\lambda )$ ($a=1,2,3$, $m\in {\mathbf{Z}}$) by
\begin{eqnarray}
{\mathcal{T}}_{1,m}(\lambda )
&=&\langle s_{-2},s_{-1},s_{0},s_{m+1}
\rangle^{[-m]},
\\
\mathcal{T}_{
2,m}(\lambda )
&=&\langle s_{-2},s_{-1},s_{1},s_{m+1}
\rangle^{[-m]},
\\
\mathcal{T}_{3,m}(\lambda )
&=&\langle s_{-2},s_{0},s_{1},s_{m+1}
\rangle^{[-m]},
\end{eqnarray}
where $f^{[m]}(\lambda )\equiv f(\lambda +\frac{m}{2}\frac{2\pi i}{4M})$.
A solution $s_{k}$ ($k\in {\mathbf{Z}}$) is expanded in terms of this basis
as
\begin{align}
s_{k}
&=-\mathcal{T}^{[k]}_{1,k-2}s_{-2}+\mathcal{T}^{[k-1]}_{3,k-1}s_{-1}-
\mathcal{T}^{[k-1]}_{2,k-1}s_{0}
+\mathcal{T}^{[k-1]}_{1,k-1}s_{1}.
\end{align}
The coefficients of $s_{-1}$, $s_{0}$ and $s_{1}$ follow from the
definition of $\mathcal{T}_{a,m}$ directly. The coefficient of
$s_{-2}$ is evaluated as $ \langle s_{k}, s_{-1}, s_{0}, s_{1}\rangle
$. Using the identity:
\begin{align}
\langle s_{i_{1}}, s_{i_{2}}, s_{i_{3}}, s_{i_{4}}\rangle^{[2]}
&=
\langle s_{i_{1}+1}, s_{i_{2}+1}, s_{i_{3}+1}, s_{i_{4}+1}\rangle ,
\end{align}
which follows from the Symanzik rotation, it is shown to be equal to
$\langle s_{-2}, s_{-1}, s_{0}, s_{k-1}\rangle^{[2]}=-\mathcal{T}^{[k]}
_{1,k-2}$.

We expand $s_{-k}$ in terms of the basis $\mathcal{X}^{(2)}_{i}$:
\begin{align}
s_{-k}(z,\bar{z})=\sum_{i=1}^{4}Q_{i}(\lambda -k \frac{2\pi i}{4M},g)
\mathcal{X}^{(2)}_{i}(z,\bar{z}|\lambda ,g),
\end{align}
where $Q_{i}:=Q_{i}^{(2)}$. The exterior product $s_{-i_{1}}\wedge s
_{-i_{2}}\cdots \wedge s_{-i_{p}}$ in $\wedge^{p} V^{(2)}$ is also
expanded in the basis $\mathcal{X}_{i}^{(2)}$. The coefficient of the
highest weight vector is evaluated as
\begin{align}
s_{-i_{1}}\wedge s_{-i_{2}}\cdots \wedge
s_{-i_{p}}=W^{(p)}_{i_{1} i
_{2} \ldots i_{p}}{\mathcal{X}}^{(2)}_{1}\wedge \cdots
\wedge {\mathcal{X}}^{(2)}
_{p}+\cdots ,
\end{align}
where we introduce the determinant
\begin{align}
W^{(p)}_{i_{1}i_{2}\ldots i_{p}}
:=
\det
\begin{pmatrix}
Q_{1}(\lambda -i_{1}\frac{2\pi i}{4M}) & Q_{1}(\lambda -i_{2}\frac{2
\pi i}{4M}) &
\cdots & Q_{1}(\lambda -i_{p}\frac{2\pi i}{4M})
\\
\noalign{\vspace{3pt}}
Q_{2}(\lambda -i_{1}\frac{2\pi i}{4M}) & Q_{2}(\lambda -i_{2}\frac{2
\pi i}{4M}) &
\cdots & Q_{2}(\lambda -i_{p}\frac{2\pi i}{4M})
\\
\vdots & \vdots & & \vdots
\\
Q_{p}(\lambda -i_{1}\frac{2\pi i}{4M}) & Q_{p}(\lambda -i_{2}\frac{2
\pi i}{4M}) &
\cdots & Q_{p}(\lambda -i_{p}\frac{2\pi i}{4M})
\end{pmatrix}
.
\label{eq:minor1}
\end{align}
For $p=1$ we have $W^{(1)}_{k}=Q_{1}^{(2)[-2k]}$. For $p=4$, we obtain
$W_{i_{1}i_{2}i_{3}i_{4}}^{(4)}=\langle s_{-i_{1}},s_{-i_{2}}, s_{-i
_{3}},s_{-i_{4}}\rangle $. In particular, from the normalization
condition {(\ref{eq:normalization1})} we find that
\begin{align}
W^{(4)}_{k-1,k, k+1,k+2}=1.
\label{eq:qw1}
\end{align}
The relation {(\ref{eq:qw1})} can be regarded as the quantum Wronskian
relation \cite{Lukyanov:2010rn}. Let us consider two more examples.
For $p=2$ with $i_{1}=-k$ and $i_{2}=-k+1$, using the $\psi $-system
{(\ref{eq:psi-system2})}, we find
\begin{align}
W_{k,k-1}^{(2)}=Q_{1}^{(1)}(\lambda -k\frac{2\pi i}{4M}).
\label{eq:w201}
\end{align}
For $W_{k+1,k,k-1}^{(3)}$, using {(\ref{eq:barsk2})}, we have
\begin{align}
\langle \bar{s}_{-k}, \mathcal{X}^{(2)}_{4}\rangle =-\frac{1}{16}W_{k+1,k,k-1}
^{(3)},
\end{align}
which becomes $\langle s_{-k}, F^{T}{\mathcal{X}}_{4}^{(2)}\rangle $ by the
formula {(\ref{eq:charge1})}. We then get
\begin{align}
W_{k+1,k,k-1}^{(3)}=16 Q_{1}^{(2)}(\lambda -k\frac{2\pi i}{4M}).
\label{eq:w3m101}
\end{align}

We note that the determinants {(\ref{eq:minor1})} satisfy the Pl\"{u}cker
relations
\begin{align}
W_{i_{0} i_{2} \cdots i_{p-1}}^{(p-1)}W_{i_{1} i_{2} \cdots i_{p}}
^{(p)}
-W_{i_{1} i_{2} \cdots i_{p-1}}^{(p-1)}W_{i_{0} i_{2} \cdots i
_{p}}^{(p)}+W_{i_{2} \cdots i_{p-1} i_{p}}^{(p-1)}W_{i_{0} i_{1}
\cdots i_{p-1}}^{(p)}=0.
\label{eq:plucker1}
\end{align}
In particular one finds
\begin{eqnarray}
0
&=&W_{0}^{(1)}W_{12}^{(2)}-W_{1}^{(1)}W_{02}^{(2)}+W_{2}^{(1)}W_{01}
^{(2)},
\nonumber
\\
0
&=&W_{02}^{(2)}W_{123}^{(3)}
-W_{12}^{(2)}W_{023}^{(3)}-W_{32}^{(2)}W
_{012}^{(3)},
\nonumber
\\
0
&=&W_{023}^{(3)}W_{1234}^{(4)}
-W_{123}^{(3)}W_{0234}^{(4)}+W_{423}
^{(3)}W_{0123}^{(4)}.
\end{eqnarray}
From these equations, we can solve $W_{0234}^{(4)}$ as
\begin{align}
W_{0234}^{(4)}=\frac{W_{0}^{(1)}}{W_{1}^{(1)}}
+\frac{W_{2}^{(1)}W
_{01}^{(2)}}{W_{1}^{(1)}W_{12}^{(2)}}
+\frac{W_{23}^{(2)}W_{012}^{(3)}}{W
_{123}^{(3)}W_{12}^{(2)}}
+\frac{W_{234}^{(3)}}{W_{123}^{(3)}}.
\label{eq:t-q1}
\end{align}
This equation is the T--Q relation of the $A_{3}$-type quantum integrable
models \cite{Dorey:2000ma}. Now using {(\ref{eq:w201})} and
{(\ref{eq:w3m101})}, {(\ref{eq:t-q1})} becomes
\begin{align}
{\mathcal{T}}_{1,1}^{[-1]}Q_{1}^{(2)}{Q}_{1}^{(1)[-1]}Q_{1}^{(2)[-2]}
&=Q
_{1}^{(2)[2]}{Q}_{1}^{(1)[-1]}Q_{1}^{(2)[-2]}
+Q_{1}^{(2)[-4]}{Q}_{1}
^{(1)[1]}Q_{1}^{(2)[-2]}
\nonumber
\\
&\quad {}
+{Q}_{1}^{(1)[-3]}Q_{1}^{(2)}Q_{1}^{(2)}
+Q_{1}^{(2)[-4]}Q_{1}^{(2)}{Q}_{1}^{(1)[-1]}.
\label{eq:t-qrel1}
\end{align}
This is the T--Q relation for the $A_{3}/{\mathbf{Z}}_{2}$-type. From this
relation we obtain the Bethe equations, which was also derived in the
previous section by using the $\psi $-system.

One can also derive a set of the relations:
\begin{eqnarray}
0
&=&W_{013}^{(3)}W_{2134}^{(4)}-W_{213}^{(3)}W_{0134}^{(4)}+W_{134}
^{(3)}W_{0213}^{(4)},
\nonumber
\\
0
&=&-W_{12}^{(2)}W_{013}^{(3)}+W_{01}^{(2)}W_{123}^{(3)}+W_{13}^{(2)}W
_{012}^{(3)},
\nonumber
\\
0
&=&W_{23}^{(2)}W_{134}^{(3)}-W_{13}^{(2)}W_{234}^{(3)}-W_{ 34}^{(2)}W
_{123}^{(3)},
\nonumber
\\
0
&=&W_{2}^{(1)}W_{13}^{(2)}-W_{1}^{(1)}W_{23}^{(2)}-W_{ 3}^{(1)}W_{12}
^{(2)},
\nonumber
\\
0
&=&W_{1}^{(1)}W_{23}^{(2)}-W_{2}^{(1)}W_{13}^{(2)}+W_{ 3}^{(1)}W_{12}
^{(2)}.
\end{eqnarray}
From these equations $W_{0134}^{(4)}$ is solved as
\begin{eqnarray}
\label{eq:T-Q-2}
W_{0134}^{(4)}
&=&\frac{W_{01}^{(2)}}{W_{12}^{(2)}}+\frac{W_{012}^{(3)}}{W
_{123}^{(3)}}(\frac{W_{1}^{(1)}W_{23}^{(2)}}{W_{2}^{(1)}W_{12}^{(2)}}+\frac{W
_{ 3}^{(1)}}{W_{2}^{(1)}})+\frac{W_{234}^{(3)}}{W_{123}^{(3)}}(\frac{W
_{1}^{(1)}}{W_{2}^{(1)}}+\frac{W_{ 3}^{(1)}W_{12}^{(2)}}{W_{23}^{(2)}W
_{2}^{(1)}})+\frac{W_{ 34}^{(2)}}{W_{23}^{(2)}}.
\end{eqnarray}
From {(\ref{eq:w201})}, {(\ref{eq:w3m101})}, {(\ref{eq:T-Q-2})} and
$W_{0134}^{(4)}=\mathcal{T}_{2,1}^{[-3]}$, we get the T--Q relation for
$\mathcal{T}_{2,1}$, $Q_{1}^{(1)}$ and $Q_{1}^{(2)}$:
\begin{eqnarray}
{\mathcal{T}}_{2,1}^{[1]}Q_{1}^{(1)[-1]}Q_{1}^{(1)[1]}(Q_{1}^{(2)})^{2}
&=&(Q
_{1}^{(2)})^{2}\left( Q_{1}^{(1)[-1]}Q_{1}^{(1)[3]} +Q_{1}^{(1)[-3]}Q
_{1}^{(1)[1]}\right)
\nonumber
\\
&&{} +\left( Q_{1}^{(2)[2]}Q_{1}^{(1)[-1]}+Q_{1}^{(2)[-2]}Q_{1}^{(1)[1]}\right)
^{2}
\label{eq:t-q12}
\end{eqnarray}
At the zeros $\lambda^{(2)}_{1n}$ of $Q_{1}^{(2)}(\lambda )$,
$\mathcal{T}_{2,1}^{[1]}$ might have a double pole. Absence of the double
pole in $\mathcal{T}^{[1]}_{2,1}$ leads to eq.~{(\ref{eq:bae1b})}. By the
shift of $\lambda $ and evaluating {(\ref{eq:t-q12})} at zeros of
$Q_{1}^{(1)}(\lambda )$, we obtain eq.~{(\ref{eq:bae1a})}. Thus one
obtains the Bethe ansatz equations again.

\section{T-system and Y-system}
\label{sec5}
Now we study the functional relations which are satisfied by
$\mathcal{T}_{a,m}$. First we calculate the product of $\mathcal{T}_{a,1}$ and
$\mathcal{T}_{1,m}$. From the Pl\"{u}cker relation
\begin{align}
& \langle s_{j_{1}}, s_{j_{2}}, s_{j_{3}}, s_{j_{4}}
\rangle
\langle s
_{i_{1}}, s_{i_{2}}, s_{i_{3}}, s_{i_{4}}
\rangle
-\langle s_{i_{1}},s
_{j_{2}}, s_{j_{3}}, s_{j_{4}}
\rangle
\langle s_{j_{1}}, s_{i_{2}}, s
_{i_{3}}, s_{i_{4}}
\rangle
\nonumber
\\
&\quad {}+\langle s_{i_{4}},s_{j_{2}}, s_{j_{3}}, s_{j_{4}}
\rangle
\langle s
_{j_{1}}, s_{i_{2}}, s_{i_{3}}, s_{i_{1}}
\rangle =0.
\label{eq:plucker2}
\end{align}
we get the identities
\begin{eqnarray}
{\mathcal{T}}_{1,1}^{[+1]}{\mathcal{T}}_{1,m-1}^{[m+1]}
&=&\mathcal{T}_{1,m}^{[m]}+
\mathcal{T}_{2,m-1}^{[m+1]},
\nonumber
\\
\mathcal{T}_{2,1}^{[+1]}{\mathcal{T}}_{1,m-1}^{[m+1]}
&=&\mathcal{T}_{2,m}^{[m]}+
\mathcal{T}_{3,m-1}^{[m+1]},
\nonumber
\\
\mathcal{T}_{3,1}^{[+1]}{\mathcal{T}}_{1,m-1}^{[m+1]}
&=&\mathcal{T}_{3,m}^{[m]}+
\mathcal{T}_{1,m-2}^{[m+2]}.
\label{eq:T-T_A_3--1}
\end{eqnarray}
These relations are a generalization of the fusion relation of the
modified sinh-Gordon equation \cite{Lukyanov:2010rn} to the modified
$B_{2}^{(1)}$ affine Toda field equation. However, one finds
\begin{eqnarray}
\label{eq:T-T_A_3-0}
{\mathcal{T}}_{1,m}^{[+1]}{\mathcal{T}}_{1,m}^{[-1]}
&=&\mathcal{T}_{1,m+1}{\mathcal{T}}
_{1,m-1}+\langle s_{-1},s_{0},s_{m+1},s_{m+2}\rangle^{[-m-1]},
\end{eqnarray}
where the second term in the r.h.s. is not the form of the $\mathcal{T}
_{a,m}$ functions. We add this function to a member of the T-functions
and define
\begin{eqnarray}
T_{1,m}(\lambda )
&=&\mathcal{T}_{1,m}(\lambda )=\langle
s_{-2},s_{-1},s
_{0},s_{m+1}\rangle^{[-m]},
\\
T_{2,m}(\lambda )
&=&\langle
s_{-1},s_{0},s_{m+1},s_{m+2}\rangle^{[-m-1]},
\label{eq:t-funct1}
\end{eqnarray}
for $m\in {\mathbf{Z}}$. The new function $T_{2,m}$ satisfies the identity
\begin{align}
T_{2,m}^{[+1]}T_{2,m}^{[-1]}
&=T_{2,m-1}T_{2,m+1}+\langle s_{-1}, s
_{m},s_{m+1}, s_{m+1}\rangle^{[-m]}
T_{1,m}.
\end{align}
We then introduce
\begin{align}
T_{3,m}(\lambda )
&=\langle s_{-1},s_{m},s_{m+1},s_{m+2}\rangle^{[-m]}.
\end{align}
But this is not new. Using {(\ref{eq:charge1})} and {(\ref{eq:barsk2})}, we
can show that $T_{3,m}=T_{1,m}$. Finally we obtain the T-system of
$A_{3}/{\mathbf{Z}}_{2}$ type:
\begin{align}
T_{1,m}^{[+1]}T_{1,m}^{[-1]}
&=T_{1,m-1}T_{1,m+1}+T_{2,m}
\nonumber
\\
T_{2,m}^{[+1]}T_{2,m}^{[-1]}
&=T_{2,m+1}T_{2,m-1}+T_{1,m}T_{1,m},
\label{eq:t-system1}
\end{align}
which is obtained by the reduction of $A_{3}$ T-system with the
identification $T_{1,m}=T_{3,m}$. Other functions $\mathcal{T}_{2,m}$,
$\mathcal{T}_{3,m}$ can be expressed in terms of $T_{a,m}$ by using
$\mathcal{T}_{2,1}=T_{2,1}^{[-1]}$, $\mathcal{T}_{3,1}=T_{3,1}^{[-2]}$ and
{(\ref{eq:T-T_A_3--1})}. They also satisfy the identities:
\begin{align}
{\mathcal{T}}_{3,m+1}{\mathcal{T}}_{1,m-1}
&=
\mathcal{T}_{3,m}^{[-1]}{\mathcal{T}}_{1,m}
^{[2]}-T_{2,m},
\nonumber
\\
\mathcal{T}_{2,m+1}^{[+1]}{\mathcal{T}}_{1,m}
&=
\mathcal{T}_{1,m+1}^{[+1]}{\mathcal{T}}_{2,m}+T_{2,m+1}.
\label{eq:t-rec1}
\end{align}

We next introduce the Y-functions by
\begin{eqnarray}
Y_{1,m}
&=&\frac{T_{1,m+1}T_{1,m-1}}{T_{2,m}},
\quad Y_{2,m}=\frac{T
_{2,m+1}T_{2,m-1}}{T_{1,m}T_{1,m}}.
\label{eq:y-system1}
\end{eqnarray}
They satisfy the Y-system of $A_{3}/{\mathbf{Z}}_{2}$ type
\begin{eqnarray}
\frac{Y^{[+1]}_{a,m}Y^{[-1]}_{a,m}}{Y_{a+1,m}Y_{a-1,m}}=\frac{(1+Y
_{a,m+1})(1+Y_{a,m-1})}{(1+Y_{a+1,m})(1+Y_{a-1,m})},
\label{eq:y-system2}
\end{eqnarray}
where $ a=1,2$ and $Y_{3,m}=Y_{1,m}$. The T-system {(\ref{eq:t-system1})}
and the Y-system {(\ref{eq:y-system2})} imply that the Langlands duality
between the modified $B_{2}^{(1)}$ affine Toda equation and the
functional equations of the $A_{3}/{\mathbf{Z}}_{2}$ quantum integrable
system.

We now discuss the boundary condition of the T-system and the Y-system.
It is easy to see that $T_{a,-1}=0$ and $T_{a,0}=1$. In order to
determine the boundary conditions $T_{1,m}$ for large $m$, we need to
study the small solutions $s_{m}$ in the whole complex plane. When
$4(M+1)$ is not a rational number, the Stokes sectors cover the complex
plane infinitely many times. So the T-functions $T_{a,m}$ are defined
independently for arbitrary positive integer $m$.

In this paper we will consider the case $4(M+1)=n$ with $n\geq 6$ being
a positive integer in detail.\footnote{When $n$ is a rational number,
we can do similar arguments. But it is not discussed in this paper.}
In this case there are $n$ Stokes sectors in the complex plane. When we
go around the origin, the solution $s_{k}(z e^{-2\pi i})$ is defined in
the sector $\mathcal{S}_{k+n}$, which is the same as $\mathcal{S}_{k}$. Then
the small solution $s_{k+n}(z)$ is proportional to $s_{k}(z e^{-2
\pi i})$:
\begin{align}
s_{k+n}(z)\propto s_{k}(z e^{-2\pi i}) .
\end{align}
For $g=0$, the linear system has no simple pole at the origin. The
solution has no monodromy around it. Then we have $ s_{k}(z e^{-2
\pi i})=s_{k}(z)$, which implies
\begin{align}
s_{k+n}(z,\lambda )\propto s_{k}(z,\lambda ).
\label{eq:monodromy0}
\end{align}
The condition {(\ref{eq:monodromy0})} leads to the boundary conditions for
the T-/Y-functions: $T_{a,n-3}=0$ and $Y_{a,n-4}=0$. The truncated
T-/Y-system becomes the same as the one for the $n$-point gluon
scattering amplitudes in AdS$_{4}$ at strong coupling
\cite{Alday:2010vh}.

For $g\neq 0$, the solutions of the linear system have monodromy around
the origin. We introduce a monodromy matrix $\Omega (\lambda )$ by
\begin{align}
\begin{pmatrix}
s_{1}
\\
s_{0}
\\
s_{-1}
\\
s_{-2}
\end{pmatrix}
 (z e^{-2\pi i},\lambda )
=\Omega (\lambda )
\begin{pmatrix}
s_{1}
\\
s_{0}
\\
s_{-1}
\\
s_{-2}
\end{pmatrix}
 (z,\lambda ).
\label{eq:monodromy2}
\end{align}
From the normalization condition {(\ref{eq:nor1})} we find $\mathrm{det}
\Omega (\lambda )=1$. We also introduce the proportionality factor
$B(\lambda )$ in {(\ref{eq:monodromy0})} for $k=1$ by
\begin{align}
s_{n+1}(z,\lambda )=B(\lambda )s_{1}(z e^{-2\pi i},\lambda ).
\label{eq:monodromy3}
\end{align}
Let us expand the solution $s_{0}(z,\lambda )$ in the basis
$\mathcal{X}_{i}(z,\bar{z}|\lambda ,g)$ whose coefficient has been defined
as $Q_{i}(\lambda ,g)$. Then we substitute its Symanzik rotation into
{(\ref{eq:monodromy3})}. In the basis $\mathcal{X}_{i}$, the monodromy matrix
becomes diagonal and takes the form ${diag}(e^{2\pi i \beta g\cdot h
^{(2)}_{1}},\ldots ,e^{2\pi i \beta g\cdot h^{(2)}_{4}} )$. Moreover
from the quasi-periodicity condition {(\ref{eq:period1})} one finds that
$B(\lambda )=-1$. Plugging {(\ref{eq:monodromy3})} into
{(\ref{eq:monodromy2})}, we get the relation
\begin{align}
\begin{pmatrix}
s_{n+1}
\\
s_{n}
\\
s_{n-1}
\\
s_{n-2}
\end{pmatrix}
 (z,\lambda )
=-\Omega (\lambda )
\begin{pmatrix}
s_{1}
\\
s_{0}
\\
s_{-1}
\\
s_{-2}
\end{pmatrix}
 (z,\lambda ),
\end{align}
which generalizes the condition {(\ref{eq:monodromy0})} and determines the
boundary condition for the T-system. It is convenient to use the
(multi-)trace of the monodromy matrix $\Omega $: $\mathrm{tr}\Omega $ and
$\mathrm{tr}^{(2)}\Omega \equiv \frac{1}{2}((\mathrm{tr}\Omega )^{2}-\mathrm{tr}
\Omega^{2})$, which are basis independent quantities. These traces can
be also expressed using the Wronskians:
\begin{align}
{\mathrm{tr}}\Omega
&=-\langle s_{-2}, s_{-1}, s_{0}, s_{n+1}\rangle
+
\langle s_{-2}, s_{-1}, s_{1}, s_{n}\rangle
-\langle s_{-2}, s_{0}, s
_{1},s_{n-1}\rangle\nonumber\\
&\quad {}
+\langle s_{-1}, s_{0}, s_{1}, s_{n-2}\rangle ,
\nonumber
\\
\mathrm{tr}^{(2)}\Omega
&=
\langle s_{-2}, s_{-1}, s_{n}, s_{n+1}\rangle
+\langle s_{0}, s_{1}, s_{n-2}, s_{n-1}\rangle
+\langle s_{-2}, s_{n-1},
s_{0}, s_{n+1}\rangle
\nonumber
\\
&\quad {}
+\langle s_{n-2},s_{-1}, s_{0}, s_{n+1}\rangle
+\langle s_{-2}, s_{n-1}, s_{n}, s_{1}\rangle
+\langle s_{n-2}, s
_{-1}, s_{n}, s_{1}\rangle .
\label{eq:monodromy4}
\end{align}
Here the r.h.s. of these equations are expressed by $\mathcal{T}_{2,m}$,
$\mathcal{T}_{3,m}$ and the auxiliary T-functions $W_{1,m}$, $W_{2,m}$,
$\bar{W}_{2,m}$ defined in \cite{Gao:2013dza}, in addition to the
T-functions {(\ref{eq:t-funct1})}. In Appendix~A, we will summarize these
auxiliary T-functions and their recursion relations. In the diagonal
basis, they are evaluated as
\begin{align}
{\mathrm{tr}}\Omega
&=4\cos (\beta g_{1}\pi )\cos (\beta g_{2}\pi )
\\
\mathrm{tr}^{(2)}\Omega
&
=2+4 \cos [\beta (g_{1}-g_{2})\pi ]\cos [
\beta (g_{1}+g_{2})\pi ]
\end{align}
where $g_{i}\equiv g\cdot e_{i}$ ($i=1,2$). For $g=0$, one finds that
$\mathrm{tr} \Omega =4$ and $\mathrm{tr}^{(2)}\Omega =6$. The monodromy
conditions {(\ref{eq:monodromy4})} determine $T_{a,n}$ ($a=1,2$). Then the
T-system extends up to $m=n-1$ and the T-functions $T_{a,m}$ for
$m\geq n-1$ are determined by the T-system and the monodromy conditions.
Concerning the Y-system {(\ref{eq:y-system2})}, it also extends up to
$m=n-2$. It is convenient to introduce new Y-functions $\bar{Y}_{a}$
($a=1,2, 3$) by
\begin{align}
\bar{Y}_{1}
&=\bar{Y}_{3}=-\frac{T_{1,n-2}}{T_{2,n-1}},\quad
\bar{Y}
_{2}=\frac{T_{2,n-2}}{T_{1,n-1}T_{1,n-1}}.
\end{align}
whose functional relations are given by
\begin{align}
\frac{\bar{Y}_{a}^{[+1]} \bar{Y}_{a}^{[-1]}}{\bar{Y}_{a+1}\bar{Y}_{a-1}}
&=\frac{1+Y_{a,n-2}}{(1+Y_{a+1,n-1})(1+Y_{a-1,n-1})}.
\label{eq:y-system3}
\end{align}
The Y-system {(\ref{eq:y-system2})} for $m=n-2$ and {(\ref{eq:y-system3})}
contains $Y_{a,n-1}$ in the r.h.s. of the equations. $Y_{a,n-1}$ are
expressed as
\begin{align}
Y_{1,n-1}
&=-T_{1,n}\bar{Y}_{1}, \quad Y_{2,n-1}=T_{2,n}\bar{Y}_{2},
\label{eq:y-system4}
\end{align}
and $T_{a,n}$ are expressed in terms of the lower T-functions. For the
$n\neq 4\ell $ ($\ell =1,2,\cdots $) case, they are also expressed in
terms of the lower Y-functions by solving {(\ref{eq:y-system1})}. Then
{(\ref{eq:y-system2})} and {(\ref{eq:y-system3})} with {(\ref{eq:y-system4})}
become the closed functional relations. Note that the present T- and
Y-systems are the same as those of form factors in AdS${}_{4}$
\cite{Gao:2013dza}. However the function $p(z)$ has different pole
structure from the present one.

In the case of even $n$ and $g_{1}=0$ (or $g_{2}=0$), one can consider
the limit to the modified sinh-Gordon equation
\cite{Lukyanov:2010rn} (or gluon scattering amplitudes in AdS$_{3}$
\cite{Alday:2010vh}), where in this limit the $SO(5)$ spinor is
decomposed into left and right-handed spinors. In this reduction the
T-functions $T_{a,m}$ reduce to the functions $T_{k}$ ($k=1,\cdots ,
\frac{n}{2}-2$), which are defined by the inner product of the left-handed
spinors. They satisfy
\begin{align}
&T_{1,2k+1}=0,\quad T_{1,2k}=-T_{k}^{[2]},\quad
T_{2,2k}=T_{k}^{[3]}T
_{k}^{[+1]},\quad T_{2,2k+1}=-T_{k}^{[2]}T_{k+1}^{[2]},
\nonumber
\\
&\langle s_{-2},s_{0},s_{1},s_{n-1}\rangle =T_{\frac{n}{2}-2}^{[n+2]},
\quad
\langle s_{-2},s_{-1},s_{1},s_{n}\rangle =T_{\frac{n}{2}}^{[n]}.\label{eq:T-reduce}
\end{align}
Here $T_{k}$ obey the functional relations
\begin{align}
T_{k}^{[2]}T_{k}^{[-2]}
&=1+T_{k-1}T_{k+1}.
\end{align}
Using {(\ref{eq:T-reduce})}, we can rewrite $\text{tr}\Omega $ in terms
of the left-handed part and right-handed part. Decomposing these two
parts we obtain $T_{\frac{n}{2}}-T_{\frac{n}{2}-2}=2\cos \pi \beta g_{2}$,
which is the trace of monodromy in left-handed part. The Y-functions
$Y_{a,m}$ and $\bar{Y}_{a}$ reduce to $Y_{k}=T_{k+1}T_{k-1}$ ($k=1,
\cdots , n/2-2$) and $\bar{Y}=-T_{\frac{n}{2}-2}$ as
\begin{align}
Y_{1,2k}
&=0, \quad Y_{1,2k+1}=-1, \quad
Y_{2,2k+1}=\infty , \quad Y
_{2,2k}=Y_{k},
\nonumber
\\
\bar{Y}_{1}Y_{2,n-2}
&=\bar{Y}^{[2]}, \quad
\bar{Y}_{2}=\infty .
\label{eq:ads3-y1}
\end{align}
Here the Y-functions $Y_{k}$ and $\bar{Y}$ satisfy the $D_{n/2}$-type
Y-system \cite{Lukyanov:2010rn,Maldacena:2010kp}
\begin{align}
Y_{k}^{[2]}Y_{k}^{[-2]}
&=(1+Y_{k-1})(1+Y_{k+2}), \quad (k=1,\ldots ,
\frac{n}{2}-3),
\nonumber
\\
\bar{Y}^{[2]}\bar{Y}^{[-2]}
&=1+Y_{\frac{n}{2}-2},
\nonumber
\\
Y_{\frac{n}{2}-2}^{[2]}Y_{\frac{n}{2}-2}^{[-2]}
&=
(1+Y_{\frac{n}{2}-3})(1-2
\cos \pi \beta g_{2} \bar{Y}+\bar{Y}^{2} ).
\end{align}

\section{High-temperature limit of the Y-system}
\label{sec6}
In the previous section we have seen that the T-/Y-system becomes the
extended one in the presence of monodromy. The standard approach to
analyze the (extended) Y-system is to derive the Thermodynamic Bethe
ansatz (TBA) equations and investigate their free energy. The IR (or
low-temperature) limit of the TBA system are characterized by the WKB
approximation, whereas in the UV (or high-temperature) limit is
characterized by the spectral parameter independent Y-functions and the
free energy is determined by the dilog formulas
\cite{Kirillov:1987,Kuniba:1993cn,Kuniba:1993nr}. Since the present
Y-system is very complicated, we leave the detailed TBA analysis to the
subsequent paper. Instead we will study the high-temperature limit of
the Y-system and their solutions explicitly for the simplest case
$n=6$.

For $n=6$. Using the T-system, $T_{a,m}$ ($a=1,2$, $m=1,\ldots ,6$) are
solved in terms of \mbox{$T_{1,1}=x$} and $T_{2,1}=y$ using
{(\ref{eq:t-system1})}. Substituting them into the monodromy conditions
we obtain two equations for $x$ and $y$
\begin{align}
&x^{6} - 6 x^{4} (-1 + y) - 2 y (-3 + y^{2}) + 3 x^{2} (-1 - 4 y + 3 y
^{2}) +
4 \cos [\beta g_{1} \pi ] \cos [\beta g_{2} \pi ]=0,
\nonumber
\\
&
-2 x^{6} + y^{2} (-3 + y^{2})^{2} + 3 x^{4} (8 - 8 y + 3 y^{2}) -
6
x^{2} (3 - 2 y^{3} + y^{4})
\nonumber
\\
& - 4 \cos [\beta (g_{1} - g_{2}) \pi ] \cos [\beta (g_{1} + g_{2})
\pi ]=0.
\label{eq:monodromy5}
\end{align}
For $g_{1}=g_{2}=0$, \textit{i.e.} when there is no monodromy around the
origin, we find the solutions of the above algebraic equations are given
by $(x,y)=(0,-1), (0,2), (\pm 2\sqrt{3},5)$ and $(\pm \sqrt{3},2)$.
For $(x,y)=(0,-1)$, we get $Y_{1,2k-1}=-1$, $Y_{1, 2k}=0$,
$Y_{2,2k-1}=\infty $ ($k\geq 1$), $Y_{2,2}=0$ and $Y_{2,4}=-1$. This
solution corresponds to the $AdS{}_{3}$ limit of the Y-system. For
$(x,y)=(\pm \sqrt{3},2)$, we get $Y_{1,1}=\frac{1}{2}$, $Y_{2,2}=
\frac{1}{3}$ and $Y_{1,2}=Y_{2,2}=0$, which corresponds to the constant
Y-system of the 6-point amplitudes \cite{Alday:2010vh}. For other
solutions, we do not find any corresponding physical quantities.

\begin{center}
\begin{figure}[htbp]
\begin{minipage}{0.5\hsize}
 \resizebox{80mm}{!}{\includegraphics{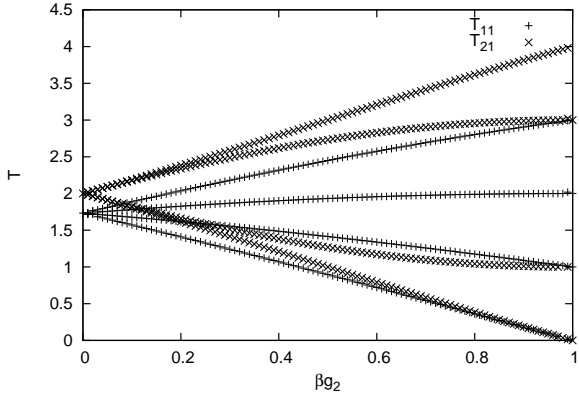}}
\vspace{1cm}
\caption{Plots of $T_{1,1}$ and $T_{2,1}$ at $g_1=0$}
\label{fig:1}
\end{minipage}
\begin{minipage}{0.5\hsize}
 \resizebox{80mm}{!}{\includegraphics{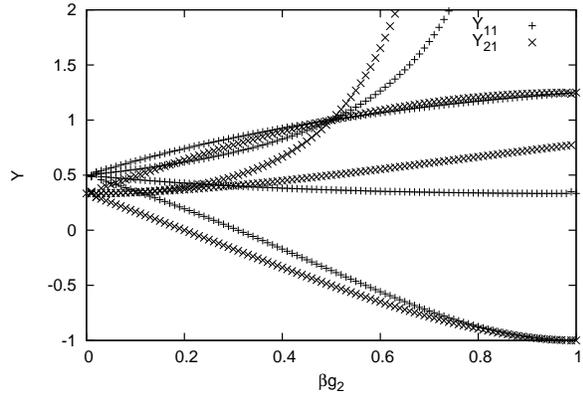}}
\vspace{1cm}
\caption{Plots of $Y_{1,1}$ and $Y_{2,1}$ at $g_1=0$}
\label{fig:2}
\end{minipage}
\end{figure}
\end{center}

Now we turn on $g_{2}$ with keeping $g_{1}=0$. We find the solutions of
{(\ref{eq:monodromy5})} with $x=0$ are given by
\begin{align}
y=\pm \sqrt{2+2\cos \frac{2\pi }{3}(1-\beta g_{2})}.
\end{align}
This gives
\begin{align}
Y_{2,2}
&=1+2\cos \frac{2\pi }{3}(1-\beta g_{2}),\quad
\bar{Y}_{1}Y_{2,4}=-y,
\end{align}
which turns out to be a constant solution. This corresponds to a
constant solution of the $AdS_{3}$ form factor with $n=6$
gluons~\cite{Maldacena:2010kp}. Note that $\bar{Y}^{[2]}=\bar{Y}_{1}Y
_{2,4}$ in {(\ref{eq:ads3-y1})} for $n=6$. As for the solutions starting
from $(\sqrt{3}, 2)$, we solve eqs.~{(\ref{eq:monodromy5})} numerically.
The graphs of $T_{1,1}=x(g_{2})$ and $T_{2,1}=y(g_{2})$ are shown in
{Fig.~\ref{fig:1}} and the corresponding Y-functions $Y_{1,1}$ and
$Y_{2,1}$ are shown in {Fig.~\ref{fig:2}}. There arise four branches from
the point $(\sqrt{3},2)$, which arrive at the solutions $(x,y)=(3,4)$,
$(2,3)$, $(1,0)$ and $(0,1)$ at $g_{2}=1$. These solutions provide a
deformation of the constant T-/Y-systems by the monodromy parameter
$g_{2}$.

\section{Conclusions and discussion}
\label{sec7}
In this paper, we studied the massive ODE/IM correspondence for modified
$B_{2}^{(1)}$ affine Toda field equation. By investigating the solutions
of the linear problem associated with the modified affine Toda equation,
we derived the $\psi $-system. This leads to the Bethe ansatz equations
corresponding to the integrable model which is not identified yet. We
also derived the same Bethe ansatz equations from the T--Q relations. We
constructed the T-system and Y-system from the Wronskians of the
solutions of the linear problem. These systems have non-trivial boundary
conditions due to the presence of monodromy around the origin. It would
be interesting to generalize the present approach to modified affine
Toda field equations associated with other affine Lie algebras which are
not of the Langlands dual of an untwisted affine Lie
algebra~\cite{Locke}. It is also interesting to study the massless
limit of this linear problem and investigate the description by using
the free field realization of conformal field theory
\cite{Bazhanov:1994ft,bazhanov1997integrable,Bazhanov:2001xm,Kojima:2008zza}.

For the linear system associated the null-polygonal minimal surface in
AdS$_{4}$, we have seen that the corresponding integrable system is the
homogeneous sine-Gordon model \cite{Hatsuda:2010cc,Hatsuda:2011ke}.
When the solution has monodromy around the origin, we have seen the
T-system and Y-system are extended and they take the form that appears
in the strong-coupling limit of the form factor in $\mathcal{N}=4$ super
Yang--Mills theory. For a general polynomial $p(z)$ and the appropriate
boundary conditions for the solutions of the linear problem, one can
describe the minimal surface problem using the massive ODE/IM
correspondence. In particular it is interesting to explore the ODE/IM
correspondence for the minimal surface in $AdS_{5}$, where the
corresponding quantum integrable model is not known yet.

\subsection*{Acknowledgments}
We would like to thank Yuji Satoh and Christopher Locke for valuable
discussion and collaboration in an early stage of this work. This work
is supported in part by Grant-in-Aid for Scientific Research 15K05043
from Japan Society for the Promotion of Science (JSPS).

\appendix{Auxiliary T-functions}
\label{appA}
In this appendix we summarize the auxiliary T-functions and their
recursion relations \cite{Gao:2013dza}. From these relations we
express $\mathrm{tr}\Omega $ and $\mathrm{tr}^{(2)}\Omega $ in terms of
$T_{a,m}$ ($m\leq n$). Furthermore we can express $Y_{a,n-1}$ in the
lower Y-functions and get a closed Y-system. We define the functions
$U_{1,m}$, $V_{1,m}$, $W_{1,m}$, $W_{2,m}$ $\bar{W}_{2,m}$ ($m\in
{\mathbf{Z}}$) by
\begin{align}
U_{1,m}
&=\langle s_{-2}, s_{-1}, s_{m},s_{m+2}\rangle^{[-m]},
\label{eq:U-1}\quad
V_{1,m}=\langle s_{-2}, s_{0}, s_{m+1}, s_{m+2}\rangle^{[-m]},
\\
W_{1,m}
&=\langle s_{-2}, s_{0}, s_{m+1}, s_{m+3}\rangle^{[-m-1]},
\\
W_{2,m}
&=\langle s_{-1}, s_{0}, s_{m+1}, s_{m+4}\rangle^{[-m-2]},
\quad
\bar{W}_{2,m}=\langle s_{-2}, s_{1}, s_{m+2}, s_{m+3}
\rangle^{[-m-2]}.
\label{eq:W-1}
\end{align}
From the Pl\"{u}cker relation {(\ref{eq:plucker2})}, we can show that
these auxiliary T-functions satisfy
\begin{align}
U_{1,m}T_{1,m}
&= T^{[-1]}_{1,m-1}T_{2,m+1}+T^{[+1]}_{1,m+1}T^{[-1]}
_{2,m},
\\
V_{1,m}T_{3,m}
&=T^{[+1]}_{1,m-1}T_{2,m+1}+T^{[-1]}_{3,m+1}T^{[+1]}
_{2,m},
\\
W_{1,m}T_{2,m}
&=V^{[-1]}_{1,m}U^{[+1]}_{1,m}-T^{[-1]}_{1,m}T^{[+1]}
_{3,m},
\\
W_{2,m}T_{2,m+1}
&= U^{[+1]}_{1,m+1}U_{1,m}-T^{[-1]}_{2,m}T^{[+1]}
_{2,m+2},
\\
\bar{W}_{2,m}T_{2,m+1}
&=
V^{[-1]}_{1,m+1}V_{1,m}-T^{[+1]}_{2,m}T^{[-1]}
_{2,m+2}.
\end{align}
Then $U_{1,m}$, $V_{1,m}$, $W_{1,m}$, $W_{2,m}$ and $\bar{W}_{2,m}$ are
expressed in terms of $T_{a,s}$. The (symmetrized) trace of the
monodromy matrix becomes
\begin{align}
{\mathrm{tr}}\Omega
&=-\left( T_{1,n}^{[n]}-\mathcal{T}_{2,n-1}^{[n-1]}
+\mathcal{T}_{3,n-2}^{[n-2]}-T_{1,n-4}^{[n-2]}\right) ,
\label{eq:tr-1}
\\
\mathrm{tr}^{(2)}\Omega
& =T_{2,n}^{[n-1]}+T_{2,n-4}^{[n-1]}
-W_{1,n-2}
^{[n-1]}+W_{2,n-3}^{[n-1]}
+\bar{W}_{2,n-3}^{[n-1]}-W_{1,n-4}^{[n-1]}.
\label{eq:tr-2}
\end{align}
Using $\mathcal{T}_{3,1}=T_{1,1}^{[-2]}, \mathcal{T}_{2,1}=T_{2,1}^{[-1]}$ and
the identities {(\ref{eq:T-T_A_3--1})}, we get
\begin{eqnarray}
{\mathcal{T}}_{3,n-2}
&=&
T_{1,1}^{[-n+1]}T_{1,n-3}^{[+1]}-T_{1,n-4}^{[+2]}
\label{eq:t3-1}
\\
\mathcal{T}_{2,n-1}
&=&
T_{2,1}^{[-n+1]}T_{1,n-2}^{[+1]}-\{T_{1,1}^{[-n+1]}T
_{1,n-3}^{[+1]}-T_{1,n-4}^{[+2]}\}.
\label{eq:t2-1}
\end{eqnarray}
The (symmetrized) traces are expressed in terms of $T_{a,s}$ as
\begin{eqnarray}
{\mathrm{tr}}\Omega
&=&-\left( T_{1,n}^{[n]}-T_{1,n-4}^{[n-2]}-\left( T_{2,1}
^{[-n+1]}T_{1,n-2}^{[+1]}-(T_{1,1}T_{1,n-3}^{[n]}-T_{1,n-4}^{[n+1]})\right)\right.\nonumber\\
&&\left. {}+(T
_{1,1}^{[-1]}T_{1,n-3}^{[n-1]}-T_{1,n-4}^{[n]})\right) ,
\nonumber
\\
\mathrm{tr}^{(2)}\Omega
&=&T_{2,n}^{[n-1]}+T_{2,n-4}^{[n-1]}
\nonumber
\\
&&{}-\left( -\frac{T_{3,n-2}^{[n]}T_{1,n-2}^{[n-2]}}{T_{2,n-2}^{[n-1]}}\right.\nonumber\\
&&\left. {}+\frac{1}{T
_{2,n-2}^{[n-1]}}\frac{T_{1,n-1}^{[n+1]}T_{2,n-2}^{[n-1]}+T_{2,n-1}
^{[n]}T_{1,n-3}^{[n-1]}}{T_{3,n-2}^{[n]}}\frac{T_{2,n-2}^{[n-1]}T_{1,n-1}
^{[n-3]}+T_{1,n-3}^{[n-1]}T_{2,n-1}^{[n-2]}}{T_{3,n-2}^{[n-2]}}\right)
\nonumber
\\
&&{}+\left( -\frac{T_{2,n-1}^{[n]}T_{2,n-3}^{[n-2]}}{T_{2,n-2}^{[n-1]}}\right.\nonumber\\
&&\left. {}+\frac{1}{T
_{2,n-2}^{[n-1]}}\frac{T_{1,n-3}^{[n-1]}T_{2,n-1}^{[n]}+T_{1,n-1}^{[n+1]}T
_{2,n-2}^{[n-1]}}{T_{1,n-2}^{[n]}}\frac{T_{1,n-4}^{[n-2]}T_{2,n-2}
^{[n-1]}+T_{1,n-2}^{[n]}T_{2,n-3}^{[n-2]}}{T_{1,n-3}^{[n-1]}}\right)
\nonumber
\\
&&{}+\left( -\frac{T_{2,n-3}^{[n]}T_{2,n-1}^{[n-2]}}{T_{2,n-2}^{[n-1]}}\right.\nonumber\\
&&\left. {}+\frac{1}{T
_{2,n-2}^{[n-1]}}\frac{T_{2,n-3}^{[n]}T_{1,n-2}^{[n-2]}+T_{1,n-4}^{[n]}T
_{2,n-2}^{[n-1]}}{T_{3,n-3}^{[n-1]}}\frac{T_{2,n-2}^{[n-1]}T_{1,n-1}
^{[n-3]}+T_{1,n-3}^{[n-1]}T_{2,n-1}^{[n-2]}}{T_{3,n-2}^{[n-2]}}\right)
\nonumber
\\
&&{}-\left( -\frac{T_{3,n-4}^{[n]}T_{1,n-4}^{[n-2]}}{T_{2,n-4}^{[n-1]}}\right.\nonumber\\
&&\left. {}+\frac{1}{T
_{2,n-4}^{[n-1]}}\frac{T_{1,n-3}^{[n+1]}T_{2,n-4}^{[n-1]}+T_{2,n-3}
^{[n]}T_{1,n-5}^{[n-1]}}{T_{3,n-4}^{[n]}}\frac{T_{2,n-4}^{[n-1]}T_{1,n-3}
^{[n-3]}+T_{1,n-5}^{[n-1]}T_{2,n-3}^{[n-2]}}{T_{3,n-4}^{[n-2]}}\right) .
\end{eqnarray}
From these equations we can write $T_{1,n}$ and $T_{2,n}$ in terms of
lower T-functions. In the case of $n\neq 4\ell $ with $\ell =1,2,
\cdots $, the T-functions can be further expressed in terms of the
Y-functions by solving {(\ref{eq:y-system1})}. We then obtain a closed
Y-system.
%

\end{document}